%% file: 4bdyCCscatt.tex
\documentclass[%
 reprint,
nofootinbib,
 amsmath,amssymb,
 aps,
]{revtex4-2}
\usepackage{graphicx}
\usepackage[utf8]{inputenc}
\usepackage[english]{babel}
\usepackage{amssymb,amsthm,amsmath,amstext,amsbsy,amsopn}
\usepackage[cal=boondoxo]{mathalpha}
\usepackage{subfigure}
\usepackage{mathrsfs}
\usepackage{dcolumn}
\usepackage{bm}
\usepackage[normalem]{ulem}
\usepackage{tikz}
\usepackage{tzplot}
\usepackage{nicefrac}
\usepackage{multirow}
\usepackage{hyperref}
\usepackage{calligra}

\usepackage{physics}

\renewcommand{\vec}[1]{\bm{#1}}
\newcommand{\bt}[1]{\ensuremath{B_3^{(#1)}}}
\newcommand{\bte}[1]{\ensuremath{B_4^{(#1)}}}
\newcommand{\bd}{\ensuremath{B_2}}
\newcommand{\ad}{a_{22}}
\newcommand{\aA}{a_2}
\newcommand{\at}{a_{31}}
\newcommand{\ham}{\ensuremath{\hat{H}_{\text{\tiny LO}}}}

\newcommand{\fm}[1]{\ensuremath{#1\,\text{fm}^{-2}}}
\newcommand{\asy}{\ensuremath{\hat{\mathcal{A}}}}

\newcommand{\la}[1]{\ensuremath{\lambda=#1~\text{fm}^{-2}}}

\begin{document}
\title{Scale-(in)dependence in quantum 4-body scattering}
\author{Sourav Mondal}
\email{sm206121110@iitg.ac.in}
\author{Rakshanda Goswami}
\email{r.goswami@iitg.ac.in}
\address{Department of Physics, Indian Institute of Technology Guwahati, Guwahati 781039, India}
\author{Johannes Kirscher}
\email{johannes.k@srmap.edu.in}
\address{Department of Physics, SRM University AP, Amaravati 522 240, Andhra Pradesh, India}
\author{Udit Raha}
\email{udit.raha@iitg.ac.in}
\address{Department of Physics, Indian Institute of Technology Guwahati, Guwahati 781039, India}
\begin{abstract}
We investigate the multi-channel \mbox{4-body} scattering system using regularized 2- and \mbox{3-body} contact interactions.
The analysis determines the sensitivity of bound-state energies, scattering phase shifts and cross sections on 
the cutoff parameter ($\lambda$), and the energy gaps between scattering thresholds. The latter dependency is obtained with
a 2-body scale fixed to an unnaturally large value and a floating \mbox{3-body} parameter. Specifically, we calculate
the binding energies of the shallow 3- and \mbox{4-body} states, dimer-dimer and trimer-atom scattering lengths, and 
the trimer-atom to dimer-dimer reaction rates. Employing a potential renormalized by a large 2-body scattering 
length and a \mbox{3-body} scale, we find all calculated observables to remain practically constant over the range 
\mbox{$6\,\text{fm}^{-2}<\lambda<10\,\text{fm}^{-2}$}. Divergences in scattering lengths emerge for critical \mbox{3-body} parameters
at which thresholds are degenerate. Such 
threshold effects are found to be independent of the regulator cutoff.

Furthermore, at those critical points
where the dimer-dimer and trimer-atom thresholds overlap, we predict an enhancement of the inelastic over the elastic
scattering event. Such an inversion between elastic- and rearrangement-collision probabilities indicates a strong 
sensitivity of the \mbox{4-body} reaction dynamics on the \mbox{3-body} parameter at finite 2-body scale. This phenomenon is absent 
in earlier studies which differ in the renormalization scheme.
As this discrepancy arises for all considered cutoffs, a more comprehensive 
parametrization of short-distance structure is necessary: sole cutoff variation does not reveal non-perturbative 
change in reaction rates conjectured to be due to a combined effect of the finite 2-body range and the specific 
choice for the \mbox{3-body} 
parameter.
\end{abstract}

\maketitle

\section{Introduction}
\label{sec.intro}

Remarkable progress has been made in the unified description of quantum systems comprising composites of atoms, nucleons,
or even exotic hadrons (see, e.g., Refs.~\cite{Mochizuki:2024dbf,Kievsky:2023xgy,Endo:2023rhh,Kinugawa:2024kwb}~for a biased 
selection of recent developments). A pronounced separation of scales in the 2-body sector of such composites allows for an 
expansion in terms of their leading-order (LO) universal contact interactions which represent a starting point for an 
effective field theoretical (EFT) analysis. In turn, features peculiar to a system, as parametrized {\it via} 
sub-leading order terms, paint a progressively sharper image of the short-distance structure of the pair interactions. This 
scheme of expansion in the zero-range contact terms and derivatives of increasing orders thereof commenced with the
application of Fermi's pseudo-potential approach to the quantum many-body problem~\cite{PhysRev.105.767}. When adapted to
nuclear physics, this scheme resembles a description of nuclei consistent with its underlying relativistic field 
theory\footnote{\label{fn.chi} Another approach considers systematically the existence of virtual pions in an expansion of 
the inter-nucleon potential~(see, e.g., Refs.~\cite{Epelbaum:2019kcf,Tews:2022yfb,Somasundaram:2023sup} for reviews and a 
recent development) yielding the most precise and ostensibly order-by-order converging postdiction for a wealth of 
observables (see, e.g., Ref.~\cite{Lonardoni:2017hgs}).} was advanced since its first comprehensive 
formulation~\cite{vanKolck:1998bw}. A major step in the development of the theory was the integration of the {\it Efimov effect}~\cite{Efimov:1971zz}~as a 
LO renormalization constraint. This refinement of the theory defied the na\"ive-dimensional-analysis ordering scheme of the expansion terms~\cite{Bedaque:1998kg},
and allowed for a much broader perspective on the linkage between systems composed of different numbers of particles.
The technical step of an unnatural enhancement of a momentum-independent contact interaction remains unique and has not yet been
found to emerge in any other few-body observable; even for particle numbers exceeding three. While the majority of analyses 
comprise observables correlated with bound-state wave functions, existing studies based on scattering systems involving more 
than three constituents,\footnote{Much effort is invested in obtaining the \mbox{4-body} scattering properties for high-precision 
interactions (see, e.g., Refs.~\cite{Flores:2022foz,Viviani:2022qfy,Fonseca:2017koi} for recent advances) over a relatively 
large energy range. Analyses on their short-distance sensitivity (see, e.g., Refs.~\cite{18434,Kravvaris:2020lhp}) are thus 
limited in their ability to expose \emph{potential} correlations between observables like the Tjon-~\cite{Tjon:1975sme} and 
Phillips-~\cite{PHILLIPS1968209} lines, while studies employing the simplistic contact interaction 
approach~\cite{PhysRevC.107.064001,Kirscher:2011uc}~are scarce.} in our opinion, are still inadequate to consider the contact
theory in its current formulation as a proper starting point for high-precision analysis of multi-particle reaction observables. 
Hence, the purpose of this work is to discuss results that allow for a more detailed understanding of the theory's pre/post-dictions
in a \mbox{4-body} scattering system.

More specifically, we analyze the dependence of \mbox{4-body} reactions on the threshold structure defined through the 2- and 
\mbox{3-body} subsystems because the former are prone to uncertainties at every order of the interaction expansion. Our selection of 
scales and particle statistics is such that the results obtained to be pertinent to a system of two protons and two neutrons 
(arguably the most prominent non-trivial \mbox{4-body} system) are general enough not to be limited to that system. This is possible as the nuclear and 
electromagnetic interactions yield a threshold hierarchy that is understood as a remnant of a universal structure which emerges for all interactions
with a ratio between its effective range and the length scale of a single 2-body state ($\aA$) goes to 
zero. Our work relates to all \mbox{4-body} systems which exhibit this separation of scales with an internal space (e.g., number of flavors, fermionic species,
spin orientations, \&c.) of dimension 
$\geq 4$. However, \emph{our} physical reality realizes this limit only approximately, and hence the extent by which \mbox{4-body} 
scattering and binding - the latter even for much larger-in-number systems - depend upon deviations from unitarity ($\aA\to\infty$), 
allows
us to understand the universal and unique character of nuclei when comparing them with atoms or any other system that exhibits 
the separation of scales.

Beyond testing the consistency of the theory, is it more than play to calculate the effect of floating thresholds whose 
experimental control and validation of predictions thereof is considered elusive for nuclei? It is because of the wealth
of cold-atom experiments~(for a review see, e.g., Ref.~\cite{Liu2020MolecularCF}) in which tuning of the 2-body scattering length (used here 
to quantify the largest scale) moves the energies of 3- or more-body states relative to the threshold set by the 2-body binding. Beyond the aid to
 such experiments,
their theoretical analogues predict features of quantum dynamics of more than academic interest if one considers 
options to enhance, e.g., nuclear fusion rates and chemical reaction yields. With differently renormalized interactions,
we find that it is the threshold structure which drives the bulk reaction rates and not details of the interaction
which yield those thresholds. In other words, we find that \mbox{4-body} reactions in a bosonic channel behave qualitatively
similar even if the the unitary limit is not fully taken. However, as this insensitivity also pertains
to a discrepancy with an earlier result for the ratio between the elastic and reaction cross sections, the verdict about the 
usefulness of contact theories in rearrangement collisions remains unanswered.

We proceed as follows. We present our results in Sec.~\ref{sec.res}~after being more precise in the ensuing two 
sections about the employed effective theory (cf. Sec.~\ref{sec.eft}) and a precise formulation of the problem (cf. 
Sec.~\ref{sec.prob}). In the appendices, we detail the non-standard renormalization of the interaction (cf. 
Appendix~\ref{app.reg})~and the numerical technique employed to obtain a variational solution to the scattering problem (cf. 
Appendices~\ref{app.rgm} and ~\ref{GA}).

\section{Problem formulation}
\label{sec.prob}

The scattering problem of four particles - each of which occupies a distinct quantum state of a \mbox{4-component} fermion - is 
considered in the context of a low-energy EFT within the 2-fragment approximation, namely, no 3- or \mbox{4-particle} breakup channels 
are taken into account explicitly. The particles are assumed to sustain a single, spherically symmetric 2-body bound-state (dimer) 
whose spatial extent is large compared with the interaction range. We investigate up to four Gaussian pair interactions, differing in 
their width, whose strengths are adjusted to yield this datum, i.e., the dimer binding energy \(B_2\). The scattering event of two such 
bound dimers off each other depends further on the 3- and \mbox{4-body} spectra which can be fixed at energies close to \(B_2\) by a 
single additional parameter. We adopt the canonical form of this parameter as the strength of a Gaussian-regularized genuine \mbox{3-body}
contact interaction. This strength controls the 3- and \mbox{4-body} spectra independently from the dimer system, and thereby enables us to 
explore the dependence of the \mbox{4-body} scattering problem constrained to two asymptotic-state fragments. Hence, we determine elastic 
scattering cross sections between two dimers, a trimer and an atom, and between the inelastic rearrangements, namely, 
dimer-dimer$~\rightleftarrows~$trimer-atom, as they depend on the threshold separation: \(\bd<\bt{\text{max}}\lesseqgtr 2\bd\). 
Furthermore, we investigate the sensitivity of these cross sections on the range of the 2- and \mbox{3-body} regularized contact 
interactions. Thereby, we assess the validity of the canonical\footnote{cf. Refs.~\cite{Contessi:2023yoz,Contessi:2024vae} for 
recent reformulations} SU(4) symmetric leading-order pionless EFT to capture features of nuclear fusion reactions.

In the following two sections, we briefly introduce the theoretical framework under investigation and the
employed numerical tools.

\section{Theory}
\label{sec.eft}
The foundational framework of pionless EFT (see, e.g., Refs.~\cite{vanKolck:1998bw,Braaten:2004rn,Hammer:2019poc} for reviews) for 
developing nuclear, atomic, and other approximately unitary systems, begins with a low-energy effective Hamiltonian at LO given by 
the general form:
\begin{equation}\label{eq.ham0}
\ham = \sum_i \hat{T}(r_i, m) + \sum_{\{i,j\}} \hat{V}_2(r_{ij}, \lambda) + \sum_{\{i,j,k\}} \hat{V}_3(r_{ij}, r_{ik}, \lambda).
\end{equation}
The Hamiltonian operator includes \( \hat{T}(r_i, m) \), the one-particle kinetic energy, and \( \hat{V}_2(r_{ij}, \lambda) \) and 
\( \hat{V}_3(r_{ij}, r_{ik}, \lambda) \), the 2- and 3-particle potentials, respectively. The latter depends solely on the 
relative coordinates between particles labeled $i,j$ and $k$. The parameter \( \lambda \) represents a regulator that distinguishes between
potential forms, all yielding a 2-body system near unitarity. In other words, regardless of the choice of the specific form of the 
potential (one feature parametrized with $\lambda$), the 2-body amplitude has a universal structure in the sense that it conjures a 
pole at a momentum that is small relative to the momenta needed to probe the spatial extent of the support of the potential. 

The inclusion of the \mbox{3-body} potential\footnote{Its strength $d$ (cf. Eq.~\eqref{eq.v3}) is denoted, for historical reasons, as TNI.} 
is one way to address an emergent ambiguity of the $A>2$ problem due to the choice of a 2-body potential that reduces to a zero-range 
contact interaction, characterized by a single strength tuned to the unitarity condition \( |\aA| \approx \infty \). Consequently, a 
scale must be introduced which is encoded here in the 3-particle operator. All this follows from understanding the Hamiltonian 
\( \ham \) as the LO of an EFT expansion, with the analysis restricted to particles of equal mass \( m \) (see Ref.~\cite{Hammer:2019poc} 
for a recent review). Our aim in this work is to analyze the potential \emph{usefulness} of this theory as a foundation for 
complex scattering processes that also involve rearrangement reactions.

The thus \emph{renormalized} interaction is deemed \emph{useful} if predictions for observables can be made whose $\lambda$ dependence
is perturbatively small. We aim to assess this for systems with large but still finite 2-body scale (nuclear motivation). As in any 
other renormalization scheme, the regularization (parametrized via $\lambda$) and a set of constraints need to be specified. We detail 
our choices in Appendix~\ref{app.reg}. Here, we only briefly mention that for computational ease we use a smooth Gaussian regulator
function and demand a fixed dimer spectrum with a single bound-state. It is notable that a numerical method of expanding wave functions
becomes increasingly inaccurate and numerically unstable for interaction ranges small compared with the support of the sought-after wave
function. Maintaining the balance between the probability density within the potential's reach and the larger part of the state residing 
outside in  the zero-range limit becomes impractical - to our knowledge - for all techniques, and we make no attempt to resolve this issue.
We thus limit our study to the ratio of the potential's range to the dimer size $r_\text{dim}/r_\text{int}\approx50$ corresponding to the
interval $\fm{4}\lesssim\lambda\lesssim\fm{10}$ for the Gaussian cutoffs\footnote{This interval exceeds what is typically assessed using 
effective chiral interactions~(cf. Refs.~\cite{Lynn:2017fxg,Lonardoni:2018nob}~and footnote~\ref{fn.chi}). It is also broader than those 
ranges within which the renormalization-scheme dependence of a contact theory~\cite{Bazak:2018qnu} was discovered and where features of 
larger systems were found to be stable~\cite{Dawkins:2019vcr}.} for a dimer binding energy of 
$\bd=0.5~\text{MeV}=0.0025\,\text{fm}^{-1}$.\footnote{We associate the potential range with the full width at half maximum of a Gaussian, 
$2\sqrt{\ln2/\lambda}$, and the spatial extent of the dimer by the distance at which the $S$-wave function drops to $1/2$, namely, 
$e^{-\sqrt{2m\bd}r}=\nicefrac{1}{2}$, which yields $r\approx 4.4\,$~fm with $1/{(2m)}\approx 40$~fm$^{-1}$.} For this sample of 2-body 
potentials, the \mbox{3-body} spectrum includes a single bound state with spatial extend of about $\ln2/\sqrt{m\bt{0}}\approx 0.46$~fm
\footnote{ 
Assuming a ground state with total hyper-angular momentum zero and binding energy $B$, its hyperradial wave function
\(u(\xi)\) obeys for 
\mbox{\(\xi\to\infty:~\left(-m^{-1}\partial_\xi^2-E\right)u(\xi)=0\;\curvearrowright\;u(\xi)\propto\exp[-\sqrt{mB}\,\xi]\)}
with \(u(\xi)=1/2\)~ at~ \(\xi=\ln2/\sqrt{m\bt{0}}\).
}
, i.e., a high probability of finding its constituents within a volume in which they interact.
To study systems where the trimer state is closer to the dimer thresholds, the renormalizing \mbox{3-body} strength needs to be repulsive with a 
strength large relative to that of the 2-body attraction. To avoid the computational complications associated with this combination of a 
relatively weak 2-body attraction with an extremely strong \mbox{3-body} repulsion, and to assess the impact of the relatively deep trimer states 
on the observables close to the dimer thresholds, we choose an attractive \mbox{3-body} strength. Consequently, the \mbox{3-body} ground state becomes 
more deeply bound, and an excited state appears as bound with $\bt{1}>\bd$. Whether the ground- or an excited trimer state is close to the 
dimer-dimer threshold at $2\bd$ should have nothing but perturbative effects on reactions. If so, by scattering off either state we assess a broader class of interaction potentials and thereby test whether the additional nodes in the trimer wave function have
(non-)perturbative effects on the \mbox{4-body} scattering system. This test is as much part of the renormalization-group evolution as the $\lambda$ 
variation.

Having outlined the overall bosonic behavior of the \mbox{4-nucleon} system, we briefly comment on the treatment of the internal degrees of 
freedom besides the space-time coordinates. This work assumes the single-particle nucleon states to occupy one of four internal states, 
namely, the four spin-isospin components. The formation of trimers and the emergence of a unique \mbox{3-body} scale - excluding the one that is 
used as a renormalization constraint - are understood as consequences of the breakdown of continuous-scaling symmetry characterizing the 
unitary 2-body subsystem and realized discretely in the 3- and more-body sectors of the {\it \mbox{4-component}} fermions. Moreover, due 
to the spin-isospin independence of $\ham$, all 6 possible dimer configurations comprised of two \mbox{4-component} fermions are degenerate; as 
are the 4 possible trimer configurations. As all the interactions preserve rotational symmetry, we assume (a) the dimer and a trimer ground
states to reside in totally symmetric spatial configurations, and (b) that the 2-fragment \mbox{4-body} cross sections are dominated by fragments 
moving relative to each other in $S$-waves. For an antisymmetric wave function, the accompanying spin-isospin part must also be 
antisymmetric. It thus suffices to consider one arrangement for the dimer-dimer channel, and one for the trimer-atom channel (cf. 
Eq.~\eqref{eq.fragwfkt}).

Predictions for systems of such particles are obtained here for bound-states by extremizing with respect to the variational parameters 
collected in $\vec{c}$ the Ritz functional:
\begin{equation}\label{eq.ritz}
[\vec{c}]=\mel{\phi}{(\ham-E)}{\phi}\,. 
\end{equation}
While for scattering observables, stationary solutions are found for the Kohn-Hulth\'ene functional~\cite{kohn1948variational}:
\begin{equation}\label{eq.kohn}
[\vec{c},{\bf a}]=\mel{\psi_{\text{\tiny ch}}}{(\ham-E)}{\psi_{\text{\tiny ch}}}-\frac{1}{2}\,a_{\text{\tiny ch,ch}}\,\,,
\end{equation}
The subscript ``ch'' refers to a specific channel or boundary condition for the scattering problem, and the variational parameter ${\bf a}$ 
is equivalent to the reactance matrix (see Appendix~\ref{app.rgm} for details). For both, we employ the numerical method of the so-called
{\it Refined Resonating Group Method}, supported by a genetic algorithm (cf. Appendix~\ref{GA}) that is used to optimize the width parameters
of the trial wavefunctions used to expand the ground and excited states of the dimer, trimer, and tetramer systems.

\section{Results}
\label{sec.res}
Employing the regularized, 3-parameter contact interaction in combination with the variational solution method, we obtain observables from 
the multi-channel scattering matrix (see Appendix~\ref{app.rgm}) as functions of: \mbox{{\bf(i)} the} cutoff $\lambda$, {\bf(ii)} the number of 
bound trimers and their respective energies $\bt{n}$, and {\bf(iii)} the dimer binding energy $\bd$. The nature of these dependencies and 
their implications are discussed below in order.

\begin{table}
\caption{\label{tab.spect} Cutoff ($\lambda$) dependence (in $\text{fm}^{-2}$) of 3- and \mbox{4-body} bound-state energies (in units of $\bd$)
for a range of $\bt{1}$. As $\bt{1}$ along with $\bd=0.5$~MeV serves as the renormalization condition, it is marked by $(\star)$.}
\begin{ruledtabular}
\begin{tabular}{ccr||rrr}
$\lambda$ & $(\star)\bt{1}$ & $\bt{0}$ & $\bte{0,1}$ & $\bte{0,2}$& $\bte{0,3}$ \\
\hline
\multirow{5}{*}{6} & 1.2 & 28.0   &        &        & 84.6   \\
                   & 1.3 & 38.8   &        &        & 120.6  \\
                   & 1.9 & 119.2  &        & 121.1  & 516.7  \\
                   & 6.0 & 835.2  &        & 2191.8 & 6457.8 \\
                   & 9.0 & 1196.6 & 1228.8 & 3706.4 & 9132.4 \\
\hline

\multirow{5}{*}{8} & 1.2 &  27     & &        & 84.8   \\
                   & 1.3 &  38.8   & &        & 121.4  \\
                   & 1.9 &  120.2  & &        & 473.2  \\
                   & 6.1 &  928.5  & & 2118.4 & 7139.6 \\
                   & 9.0 &  1366.0 & & 3758.6 &10488.8 \\ 
                   
\hline

 \multirow{4}{*}{10} & 1.2 & 28.8   & &        & 90.6    \\
                     & 1.3 & 37.6   & &        & 120.0   \\
                     & 6.2 & 1011.6 & & 2339.8 & 7952.4  \\
                     & 9.0 & 1510.0 & & 3924.0 & 11670.6 \\
                     
\end{tabular}
\end{ruledtabular}
\end{table}

\begin{figure*}[t]
\centering
\includegraphics[scale=0.2]{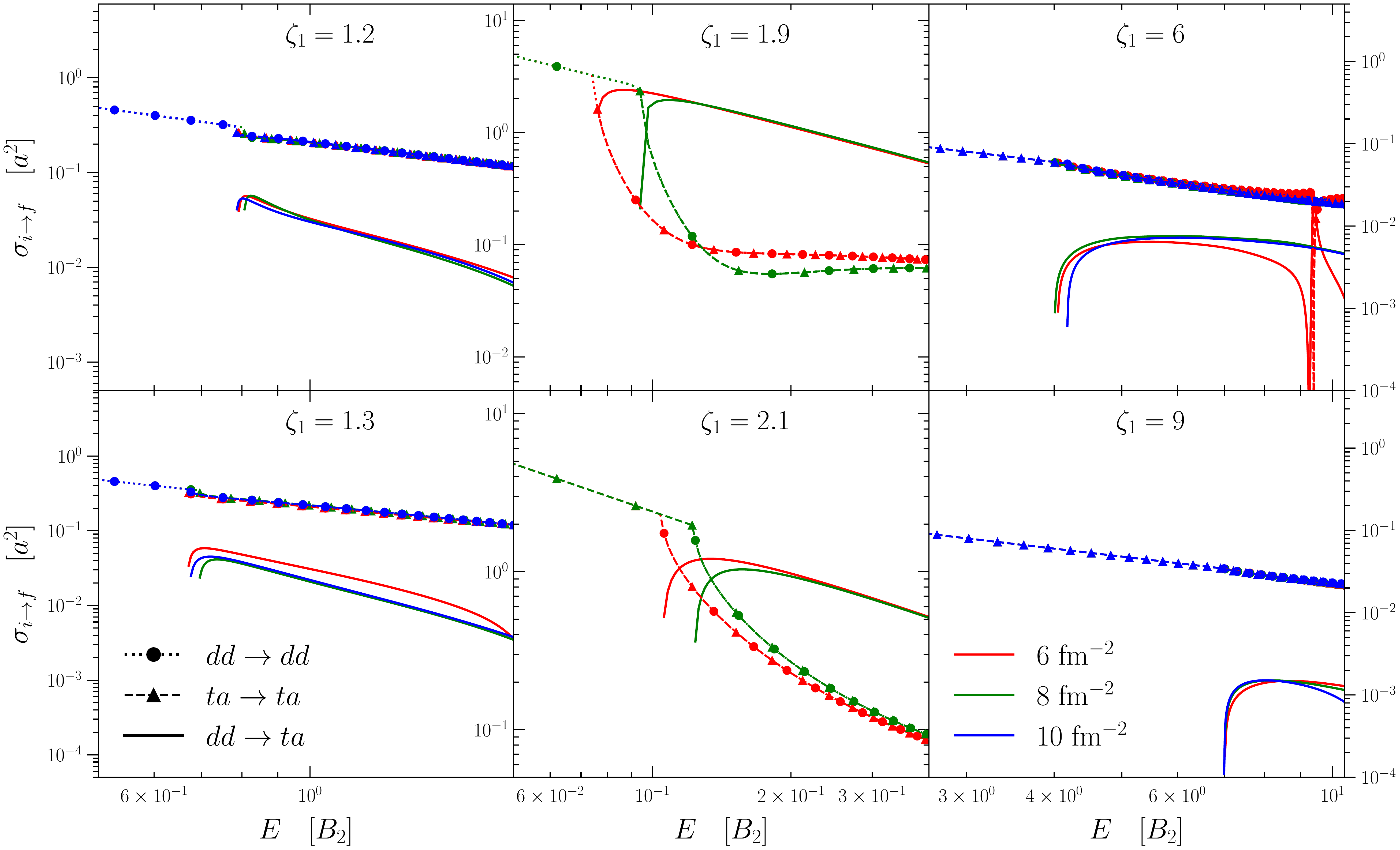}
\caption{Reaction- ($dd\to ta$,( solid lines)) and elastic ($dd\to dd$ (dotted lines) and $ta\to ta$ (dashed lines))
cross sections for a trimer binding energy below ($\bt{1}<2\,\bd$, left column), degenerate with ($\bt{1}\approx 2\,\bd$, 
middle column), and above ($\bt{1}>2\,\bd$, left column) the dimer-dimer threshold. Results for all $\zeta_1:=\bt{1}/\bd$
are shown for different regulator cutoffs \la{6}~(red), \la{8}~(green), and \la{10}~(blue), for a fixed 
$\bd=0.5~$~MeV. In each panel, the total energy $E$ is taken relative to the lowest threshold therein.}
\label{fig:crossA}
\end{figure*}

\begin{figure*}[tbp]
\centering
\includegraphics[scale=0.5]{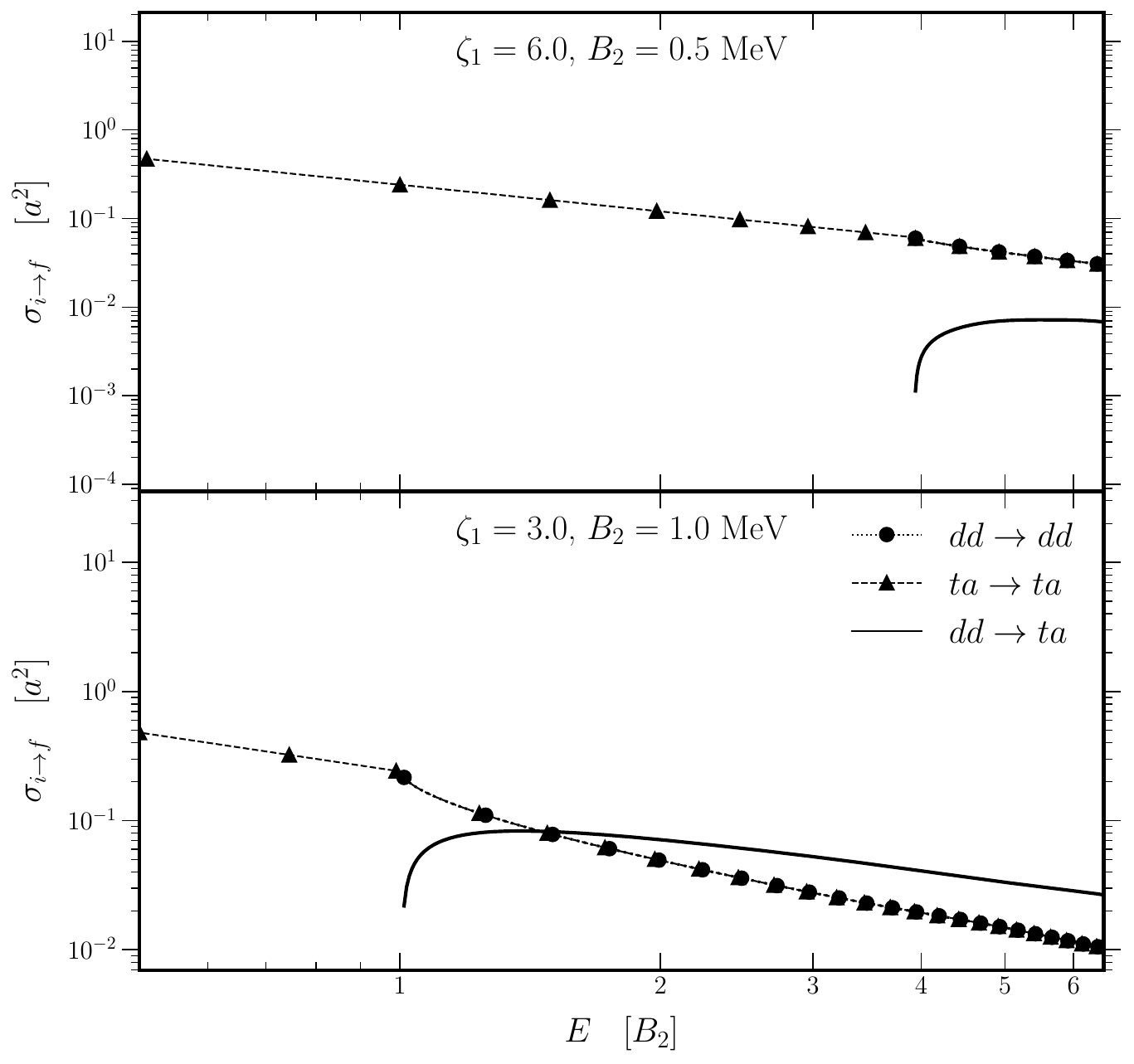}
\caption{Reaction- ($dd\to ta$ (solid lines)) and elastic ($dd\to dd$ (dotted lines) and $ta\to ta$ (dashed lines))
cross sections for a total energy $E$ relative to the $ta$ threshold set by $\bt{1}=3$~MeV. The two panels compare results
obtained with a wider ($2$~MeV, top panel) and a narrower ($1~$MeV, bottom panel) gap between trimer-atom and dimer-dimer thresholds.
}
\label{fig:crossB2}
\end{figure*}

\paragraph*{\bf (i)}
None of the observables exhibits any significant cutoff dependence within the considered interval between \fm{6} and \fm{10}. Neither 
the shallowest tetramer bound-state (cf.~Table~\ref{tab.spect}), the dimer-dimer ($\ad$) and trimer-atom ($\at$) scattering lengths 
(cf.~Table~\ref{tab.a}), nor the reaction cross sections (cf.~Figs.~\ref{fig:crossA}~and~\ref{fig:crossB2}), change significantly. This 
is noteworthy, especially for those choices of the TNI which yield diverging $\ad$ and/or $\at$. Such divergences are reflections of 
an ``old'' channel being overtaken by a ``new'' channel
\footnote{Wordings adopted from Ref.~\cite{newton2002scattering}, Chapter 17.2.2}
when the TNI is tuned such that, e.g., $\bt{x}\approx2\bd$
(see~Fig.~\ref{fig:crossA}, center column).
The signatures of a divergent scattering length in our numerical simulations is a rapidly changing magnitude and
sign as a consequence of infinitesimal variations of the basis. This rapid change is equivalent with an uncertainty of the order of the considered value,
and hence we abstain of quoting those in Tab.~\ref{tab.a}. The more rigorous wording of our result would thus report a discontinuous jump of the scattering
lengths at respective critical $\zeta$'s while we have provided strong arguments that these jumps resemble divergences.
To elaborate on the cutoff dependence of those critical points,
this constitutes nothing but a sanity check of numerics because features associated with channels opening must be cut-off independent by construction. Thus, the 
cutoff insensitivity of an ``old'' channel's amplitude is expected, but not the apparent total absence of divergences for situations when the threshold separation is 
relatively large. Even in that latter case, no non-perturbative cutoff effects were observed. We interpret this as a strong indication that our $\lambda$ variation does not
bring isolated \mbox{4-body} poles sufficiently close to the scattering thresholds because such an approach of a pole would have a significant impact on the phase shifts.
\begin{figure*}
\centering
\include{./graphs/4bdyEfimovGraph}
\caption{Dependence of the \mbox{4-body} thresholds on the 2-body scattering length $\aA$. In our analysis, we fixed
$\aA=10$~fm (vertical solid red line), and all other curves including the dimer-dimer ($dd$, black thick curve) and dimer-atom-atom 
($daa$, solid gray curve) break-up thresholds represent expected behavior for general $\aA$.
Solid lines refer to thresholds fixed by a renormalization condition while dashed ones represent hypothetical cutoff
dependence for unconstrained levels.
Four \mbox{3-body}-parameter scenarios are shown: in (I), the second excited trimer state (upper red dot) is put at the $daa$ threshold
with the accompanying first excited state at 
$E\approx -4.5$~MeV; (II) \& (III) fix the first excited trimer state between the $daa$ 
and $dd$ thresholds, and below the $dd$ threshold, respectively (orange lines); (IV) locates the first excited trimer (lower red dot)
at the $dd$ threshold.
Close-to-threshold ((I) and (IV)) and well-separated ((II) and (III)) isolated \mbox{4-body} poles are marked with a dotted line
whose colour identifies the scenario (green: (I), orange: (II), and blue (IV)). Units and scales are chosen arbitrarily
in order to display all relevant thresholds in the qualitatively correct order.}
\label{fig:4bdyEf}
\end{figure*}
\begin{figure*}
\centering
\includegraphics[scale=0.2]{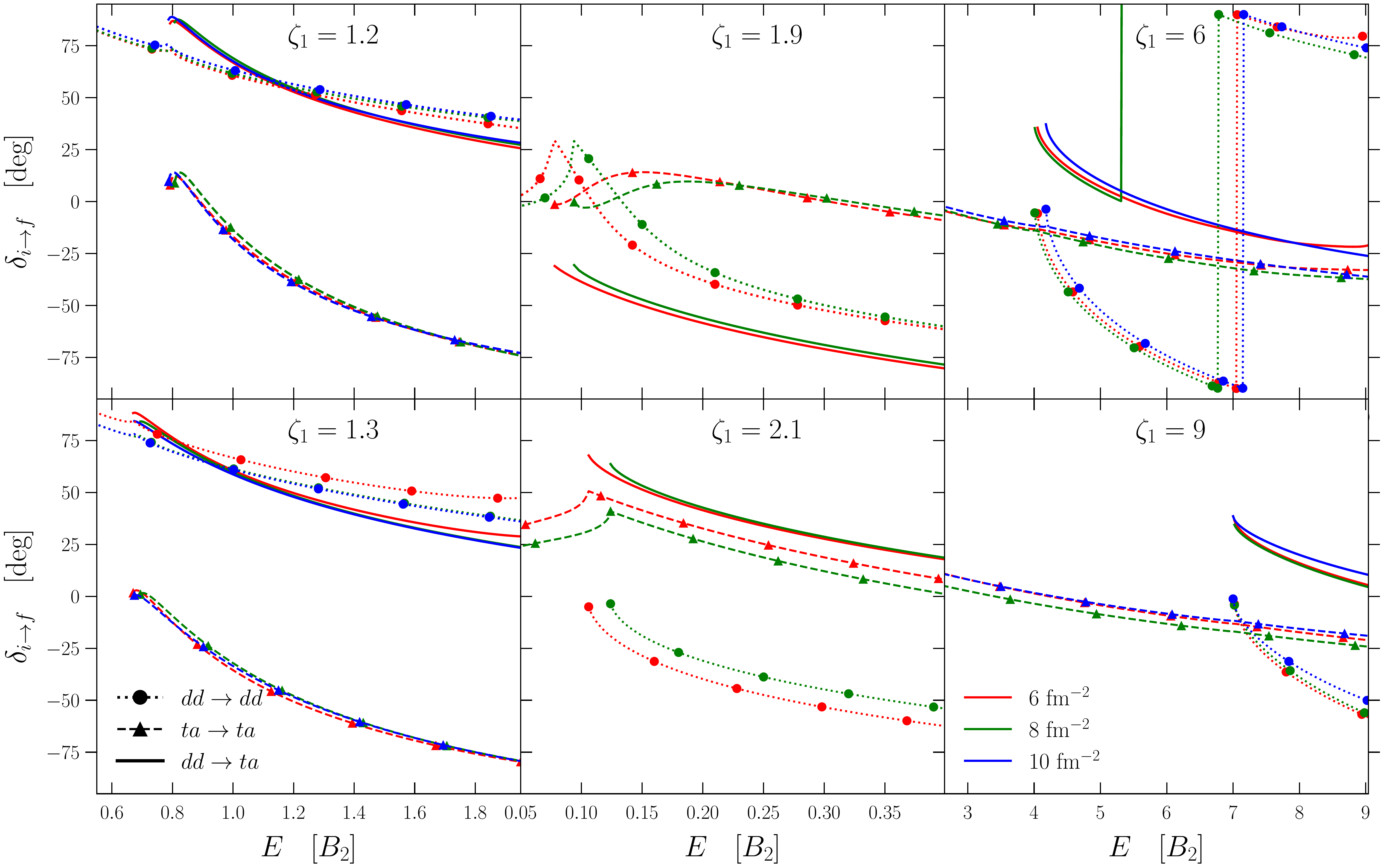}
\caption{Reaction- ($dd\to ta$ (solid lines)) and elastic ($dd\to dd$ (dotted lines) and $ta\to ta$ (dashed lines)) 
phase shifts for a trimer binding energy below ($\bt{1}<2\,\bd$, left column), degenerate with ($\bt{1}\approx 2\,\bd$, 
middle column), and above ($\bt{1}>2\,\bd$, left column) the dimer-dimer threshold.}
\label{fig:crossB}
\end{figure*}
\begin{table}[ht]
    \caption{\label{tab.a} Cutoff dependence (in $\text{fm}^{-2}$) of dimer-dimer ($\ad$) and trimer-atom ($\at$)
    scattering lengths (in fm) for a range of ratios $\bt{1}/\bd=:\zeta_1$.}
    \begin{ruledtabular}
    \begin{tabular}{cc||rr}
    $\lambda$ & $ \zeta_1 $ & $\ad$ & $\at$ \\
\hline 
\multirow{8}{*}{6}      & 1.2 & 21.0(20) & -19.0(20) \\
                        & 1.3 & 19.0(15) & -3.7(6)  \\
                        & 1.5 & 16.5(15) & 3.9(11)  \\
                        & 1.9 & 4.5(5)&  -0.8(16)\\
                        & 2.1 & 16.0(10) & -30(10)  \\
                        & 4.0 & 9.9(8)   & -7.1(9)  \\
                        & 6.0 & 7.1(2)   & -8.9(15) \\
                        & 9.0 & 4.5(2)   & -12(16)  \\
 \hline 
 \multirow{8}{*}{8}     & 1.2 & 21.6(19) & -19.7(27) \\
                        & 1.3 & 20.5(11) & 2.1(11)   \\
                        & 1.5 & 15.3(17) & 4.6(9)    \\
                        & 1.9 & 2.5(5)   & -0.3(17)  \\
                        & 2.1 & 16.0(20) & -40(20)   \\
                        & 4.0 & 10.0(5)  & -12.2(30) \\
                        & 6.1 & 6.9(2)   & -9.4(11)  \\
                        & 9.0 & 3.8(2)   & -10.1(18) \\
\hline
\multirow{5}{*}{10}     & 1.2 & 21.5(25) & -21.0(30) \\
                        & 1.3 & 20.8(17) & -1.1(14) \\
                        & 4.0 & 10.0(9)  & -9.2(20) \\
                        & 6.2 & 6.8 (2)  & -9.4(9)  \\
                        & 9.0 & 3.3(3)   & -10(15)  \\
\end{tabular}
\end{ruledtabular}
\end{table}

\paragraph*{\bf (ii)}
Next, we analyze the sensitivity of our results in regard to the gap between the trimer-atom ($ta$) thresholds and the fixed 
dimer-dimer ($dd$) and dimer-atom-atom ($daa$) thresholds. We induce the floating $ta$ threshold by smoothly changing the TNIs from 
a value that fixes the first excited trimer state to the $daa$ threshold ($\bt{1}=\bd$) up to the critical point where $\bt{1}$ goes
down below the $dd$ threshold and becomes attractive enough to sustain a second excited trimer state at the $daa$ threshold. This 
process is depicted in Fig.~\ref{fig:4bdyEf}~representing our prototype \mbox{4-body} spectrum at a distance afar from unitarity marked by
the vertical red line at $\aA=10~$fm. Here, our four choices for the TNI are 
singled out as follows: 
\begin{itemize}
\item {\bf Scenario (I):} The TNI sustains three \mbox{3-body} bound-states $t_{\rm I}^{(n=0,1,2)}$~\footnote{Notation: The floating 
threshold for the trimer states ($n=0$ stands for the ground and $n=1,2,...$ for the excited states) is denoted as 
$t_{\text{\tiny scenario}}^{(n)}$; scenario $=$I,~II,~III,~IV, with $\bt{n}/\bd=:\zeta_n$.} of which the second 
(shallowest) excited state $t_{\rm I}^{(2)}$ is close to $daa$ break-up threshold with $\zeta_{2}\approx1$. The first excited
state is, on the other hand, more strongly bound with $\zeta_{1}\approx9$, a ratio larger than in the nuclear case where the 
triton and deuteron binding energies yield $8.48~\text{MeV}/(2.22~\text{MeV})=3.82$, but still significantly smaller than the one 
presumably approached in the limit $\bd\to0$, which is at least $(22.7)^2\simeq 515.29$. The TNI is more attractive in this scenario 
compared with the other three in which only two \mbox{3-body} states are bound. 
\item {\bf Scenario (II):} The excited trimer state $t_{\rm II}^{(1)}$ lies about midway between the $daa$ and $dd$ thresholds with 
$\zeta_{1}\approx 1.5$.
\item {\bf Scenario (III):} The excited trimer state $t_{\rm III}^{(1)}$ lies below the $dd$ threshold with $\zeta_{1}\approx 6\gg2$. 
\item {\bf Scenario (IV):}  The excited trimer state $t_{\rm IV}^{(1)}$ lies at $dd$ threshold with $\zeta_{1}\approx 2$. 
\end{itemize}
When we refer to these scenarios below, the reader might find it helpful to locate those in Fig.~\ref{fig:4bdyEf}~along with the
corresponding energy levels.

In scenarios (II) and (III), we find neither $\ad$ nor $\at$ are large compared with their respective magnitudes in scenarios (I) and
(IV) (see Table~\ref{tab.a}, and note that we do not display the divergent results for $\zeta_1\approx1$ because those fluctuate between
$-\infty$ and $+\infty$ depending on otherwise insignificant changes in the variational basis).
In (I), the large $\ad$ (compared with the 2-body scattering length $\aA$) cannot be explained by the 
presence of a nearby trimer state. If the almost diverging $\ad$ would be due to $t_{\rm I}^{(2)}$, the effect of the even closer 
$t_{\rm II}^{(1)}$ should be stronger and not, as calculated, weaker (cf. $\ad$ values for
$\zeta_1=1.2$ and $\zeta_1=1.5$ in Table~\ref{tab.a}). If not a channel, i.e., branch-point effect, the large $\ad$ could be caused by an isolated pole.
This hypothesis is consistent with a universal ratio established earlier (see~\cite{Deltuva:2010xd}~where, as pertinent to our case, the
ratio was established for resonant states in the trimer-atom continuum) between an Efimov trimer and the shallowest and next-shallowest
tetramer resonance below $\bt{2}$. With $\bd\approx0.5~$MeV and only two/three bound trimers, our calculation is not in the unitary limit,
and we expect some deviation from all constants obtained in that limit. Here, either the deep or shallow tetramer resonance could be
responsible for the large $\ad$ if its energy has a ratio of $\bte{\text{2,\tiny deep/shallow}}/\bt{2}\approx 2$ (the first superscript of $B_4$
indicates the trimer it is associated with, the second starts with 1 for the shallowest and increases up to the deepest).
This ratio is significantly different from
the ratios obtained in~\cite{Deltuva:2010xd}~of $\bte{\text{n,\tiny deep}}/\bt{n}\approx 4.6$ and $\bte{\text{n,\tiny shallow}}/\bt{n}\approx 1.002$.
If we assume that nuclei emerge from the unitary limit and thus the excited and ground state of the $\alpha$ particle are tied to the universal
tetramer pair, the universal ratio for the deep state must decrease to $B_\alpha/B_{\text{\tiny triton}}\approx3.4$ while for the resonant $J^\pi=0^+$ state
one should still obtain a value close to one, $(E(0^+)+0.86~\text{MeV})/B_{\text{\tiny triton}}\approx1.06$
\footnote{We shift the energy by the expectation value for the electromagnetic
interaction as calculated in~\cite{Pudliner:1997ck}~in order to approximate the uncharged nucleus and use $E(0^+)=(28.3-20.21)~$MeV~\cite{FIARMAN19731}.}.
The former ratio is with $3.4$ still significantly larger than $2$, and thus the dependence of the \mbox{4-body} states attached to the excited trimer at $\bd=0.5\,$MeV
is not what one would extrapolate na\"ively when assuming that increasing $\bd$ from 0.5~MeV to 2.2~MeV leads to an expected decrease
of the deep tetramer's energy relative to the trimer state. However, as we are unaware of other methods,
numerical simulations, extending this one, are necessary in order to assess whether the
universal threshold positions and tetramer levels can be connected smoothly with the nuclear spectrum
by adjusting $\aA$ appropriately with the (iso)spin independent LO employed here.

In this context, it is noteworthy that for all considered trimer scenarios, the shallowest \mbox{4-body} bound-state ($\bte{n,1}$) appears with 
$1.0\lesssim\bte{0,1}/\bt{0}\lesssim4.3$~.
Assuming the presence of an isolated \mbox{4-body} pole close to $\bt{0}$ (cf. the aforementioned correlation of a universal pair of tetramers with
each Efimov trimer), we must conclude that the character of this pole changes with the
threshold separation and/or the cutoff because it appears bound only for few choices of
renormalization conditions in our calculations.
This is manifest in Tab.~\ref{tab.spect}. For $\lambda=\fm{6}$ and $\bt{1}=1.9~$MeV, the tetramers below
$\bt{0}$ assume values close to the universal ones, while for $\lambda=\fm{8}$ a tetramer with $B_4/\bt{0}\approx 1$ is absent in the spectrum and
is thus conjectured to have changed character from bound to virtual or resonant state.
Given the finding in~\cite{Deltuva:2012ig}~of a back and forth resonance- to virtual-pole transformation induced by an $\aA$ variation,
we assume the presence of such virtual and resonant poles in our spectrum, too, where the transition into those from a bound
state is induced by a change in $\lambda$ and the \mbox{3-body} parameter instead of $\aA$.
The strong dependence not only of the character of the shallow tetramer but also the variation in the deepest's binding energy
one on the renormalization scheme is not unexpected. Both energy scales are well beyond the expected range of applicability of
the LO contact theory. This range of applicability is expected to extend from $\bt{1}$ up to a breakdown energy at which the nucleon's
substructure is resolved. A conservative estimate for this energy is 
$E_{\tiny\text{break}}\approx\nicefrac{m_\pi^2}{(2\mu_{\tiny\text{max}})}\lesssim20\,\bd$ which is $\ll\bt{0}$.
Within this energy range, \mbox{4-body} poles represent resonances or virtual states which the numerical method, as used, cannot place accurately.
The fact that the tetramer bound state exhibit significant
regulator dependence compared with the scattering systems is a reflection of the energy gap between those scales. For instance,
the existence (absence) of a shallow tetramer at $\la{6}$ ($\la{8}$) for $\zeta_1=1.9$ (Tab.~\ref{tab.spect}) has little
effect on the relative smallness of the scattering lengths which attests for the detachment of the deeper part of the \mbox{4-body} spectrum
from the low-energy region.
Next, it is the number of \mbox{4-body} bound states, which we observe to increase from 1 to 2 to 3 with $\zeta$, we comment on.
This observation is contrary to the empirical conjecture~\cite{Hammer_2007}~of a universal pair of tetramer states
linked to each Efimov trimer. The suggestive resolution is to interpret the
trimer ground state as non-Efimovian whence we would not expect it to correlate with only two tetramers.
The obvious question is then, whether the number of tetramers below the deepest trimer at a critical TNI at which a shallow trimer
is degenerate with the dimer is a function of $\aA$ and thereby can be found to constitute a universal number in the unitary limit.

Analyzing the \mbox{4-body} bound states for which our numerical method yields robust predictions,
that tetramer with a binding energy closest to the
trimer ground state is expected to diverge when $\aA$ is sufficiently small, i.e., at a certain distance afar from unitarity~\cite{Ferlaino:2009zz,RevModPhys.89.035006}
\footnote{To our knowledge, an extension of the argument in~\cite{PhysRevLett.30.25}~to four particles as a more rigorous
explanation of the detachment of the deepest tetramers from the $ta$ thresholds has not been found.}.
The values in Tab.~\ref{tab.spect}~suggest that with $\aA=10~$fm, we are still far away from this regime
as the shallowest tetramers' energies do not exceed $4\cdot\bt{0}$, and it is only the deeper tetramers which seem to
detach from the trimer ground state with energies up to $8\cdot\bt{0}$ thereby becoming non-universal features along
with the appearance of a third tetramer. The latter is unexpected for true Efimov trimers higher up in the
spectrum to each of which a pair of universal tetramers is presumably linked.
Qualitatively, our result is consistent with the extended Efimov plot advanced in~\cite{Deltuva_2013}~at an $\aA$ which sets the $n$-th trimer just below $daa$,
and finds a \mbox{4-body} state close to $dd$ for the same $\aA$. An adaptation of that plot is shown in Fig.~\ref{fig:4bdyEf}, and the following discussion
is oriented around this sketch.

In our scenario (II), with a TNI less attractive compared with the one in (I), the sole excited trimer resides between thresholds (upper solid orange line)
and both $\ad$ and $\at$ keep decreasing in the course of increasing the TNI until scenario (IV) is reached. 
Prior to that, the smoothly decreasing scattering lengths do not hint towards any resonances coming close to the $dd$ or $t_{\text{\tiny x}}^{(n)}a$ thresholds.
Hence, we hypothesize those poles at locations well-separated from thresholds (upper dotted orange line).
In (IV), we obtain divergences for $\ad$ and $\at$ when $\zeta_{1}\approx 2$. 
Increasing the \mbox{3-body} attraction beyond this point, no further discontinuities are observed in either scattering amplitude 
until the interaction becomes strong enough to sustain a third timer bound-state, i.e., scenario (I).
Increasing the binding energy of this second excited state by further increasing the TNI such that is passes
the same thresholds as the first excited state when it went from scenario (II) $\to$ (IV) $\to$ (III), does
have the same effect. The difference in the number of bound-states in the \mbox{3-body} spectrum thus has no observable
effect. This is an established fact in the unitary limit, i.e., for any choice of the \mbox{3-body} parameter, observables
correlated with the Efimov spectrum exhibit discrete scale invariance. Here, we find the 2-fragment scattering matrix
for energies below the dimer-breakup energy invariant to the identification of the trimer threshold through a first or
second excited \mbox{3-body} state.

Furthermore, the results for $\ad$ indicate its foreseeable correlation with that
trimer binding energy which is closest to the dimer threshold - or any other state in the universal part of the spectrum
because it is only this part of the spectrum which is fixed to be regulator independent while we observe the deeper states to
exhibit significant cutoff and regularization-scheme dependence which would transcend into
observables like $\ad$.
Apparent in Table~\ref{tab.a}, this correlation is not linear
(cf. the Phillips correlation~\cite{PHILLIPS1968209} between the triton binding energy and
the doublet deuteron-neutron scattering length),
and $\ad$ rapidly increases whenever
the interaction is either strong enough to bind another excited trimer ($\zeta_n=1+\epsilon$)
\footnote{Throughout the text, we use $0<\epsilon\ll1$.}
or obtains a $\bt{n}$ at the $dd$ threshold ($\zeta_n=2+\epsilon$). Then, $\ad$
decreases when the trimer is moved away from either threshold.
The fact that neither $\at$ nor $\ad$ rise steeply slightly below critical $\zeta$'s
suggests the absence of resonances and the explanation of the large values as threshold
effects.

With regards to the universal ratio
obtained for \mbox{2-component-fermion} dimers~\cite{PhysRevLett.93.090404}~of $\ad/\aA\approx0.6$, this is found here for the
\mbox{4-component-fermion}, i.e., bosonic case, for trimer energies $2\,\bd\lesssim\bt{n}\lesssim6\,\bd$.
This interval includes the nuclear hierarchy and is qualitatively consistent with the
results in~\cite{Deltuva:2022zqd}.
If the trimer is bound between thresholds, $1\lesssim\zeta\lesssim2$, $\ad>a$. We abstain
from quoting results for $\ad$ in the limit $\zeta\to1$ which exhibit strong fluctuations
with variations of the variational basis which have almost no effect for other $\zeta$ scenarios.
However, a diverging $\ad$ in this limit would be perfectly consistent with the strong sensitivity
to variational parameters if it were due to the presence of the above-mentioned \mbox{4-body} pole which one
basis places very close to the $dd$ threshold while another basis shifts it further away and thereby
drastically reduces $\ad$. 

After this discussion on observables related to the elastic scattering event we turn to our findings for
the effect of the floating trimer threshold at finite and fixed 2-body scale $\bd$ on reactions.
We express results as cross sections which quantify the probability that an initial state $i$ with two asymptotically free fragments
(two dimers, an excited trimer and single particle ($ta$)) scatters into a final, asymptotic 2-fragment state $f$.
The relation to the 2-fragment scattering matrix is

\begin{multline}\label{eq.trans}
\sigma_{i \to f}(E) = \frac{\pi}{2 \mu E} \sum_{J^\pi} \frac{2J + 1}{(2s_1 + 1)(2s_2 + 1)} \\
\times \sum_{{l_i, l_f\atop s_i, s_f}} \left| S_{i\to f}\left(E,l_{i/f},s_{i/f},J\right) \right|^2\quad,
\end{multline}

with incoming-channel parameters: fragment spins \( s_{1,2} \), reduced mass \( \mu \), and fragment-relative kinetic energy \( E \).
The interaction theory predicts the all fragment's ground states with zero total angular momentum
, i.e., \( l_i, l_f = 0 \). For the dimer spin, we made the arbitrary choice \( s = 1 \). As the interaction is spin-independent, a spin singlet 
with isospin one would yield identical results. For both trimer and atom/particle/nucleon, \( s = \frac{1}{2} \). Hence, $J=S$ with
$\left[s_1\otimes s_2\right]^S$ and can assume values 0,1 for a trimer-atom channel, and 0,1,2 for the dimer-dimer case.
The interaction yields degenerate results for all \mbox{$J$ values} and does not couple channels with different $J$.
In Fig.~\ref{fig:crossA}, we show cross sections for three threshold separation choices (cf. Fig.~5 in~\cite{Deltuva:2011ur}):
$\zeta_{1}=1.2$ (left column, scenario (II) in Fig.~\ref{fig:4bdyEf}),
$\zeta_{1}=(2\pm\epsilon)$ (center column, bottom ($+\epsilon$) and top ($-\epsilon$), scenario (IV)), 
and $\zeta_{1}>6$ (right column, scenario (III)).
In scenarios (II) and (III), the transition probability is almost an order of magnitude smaller compared with the
elastic event. In both scenarios, the kink at the energy for the new-channel opening,
$\approx0.7\bd$ (left column in Fig.~\ref{fig:crossA}) with $dd$ being the old and $ta$ the new channel,
$>4\bd$ (right column in Fig.~\ref{fig:crossA}) with now $ta$ being old and $dd$ new channel, is not
shifting the power-law dependence on the energy significantly. Qualitatively, this matches the behavior discovered
earlier in~\cite{Deltuva:2011ur}.

In scenario (IV), when $\zeta_{1}=(2\pm\epsilon)$, with $\epsilon\ll1$ and hence a small threshold separation (central column, Fig.~\ref{fig:crossA}),
it is the transition cross section which assumes values as large as the elastic cross section of the old channel right before the opening of the
new one. This behavior is symmetric in the sense that it is found for the $ta$ channel opening slightly above $dd$ (upper mid-graph) and also
if it is the $dd$ channel that is higher in energy. In both cases, a rearrangement reaction is more likely compared with the incoming fragments
scattering off each other elastically.
This difference between the elastic and inelastic cross sections is peculiar and not observed in~\cite{Deltuva:2011ur}~where the respective
cross sections are almost identical. A potential explanation of this phenomenon is the differently chosen \mbox{3-body} parameter. 
The latter determines the energy of the trimer tower in the unitary limit,
and it is unclear how the changing character of the \mbox{4-body} resonances, as expected from~\cite{Deltuva:2012ig}, affects
reaction rates. A smooth variation of $\aA$ while renormalizing to the same critical $\zeta_{1}=2$ might experience
a close-to-threshold resonance instead of a close-by virtual pole.
The investigation of this issue is beyond the scope of this work as it would be incomplete without any direct
characterization of the \mbox{4-body} spectrum in the continuum. However, the following analysis offers some explanation
despite the limitations of our numerical methods.

\paragraph*{\bf (iii)}
In order to investigate this potential sensitivity with respect to certain implicit assumptions about the short-distance structure of
the 2- and \mbox{3-body} interaction which is not detected by our cutoff variation, we considered the effect of a change in the
2-body scattering length such that $\bd=1~$MeV on cross sections. To that end, we fixed $\bt{1}=3~$MeV, and compare in
Fig.~\ref{fig:crossB2}~elastic and reaction cross sections for $\zeta_{1}=6$ (upper panel) and $\zeta_{1}=3$ (lower panel).
Compared with the case already shown in the right-corner panel of Fig.~\ref{fig:crossA}, the reduction of the gap between
$dd$ and $ta$ significantly increases the reaction cross section (solid line, lower panel, Fig.~\ref{fig:crossB2}). For
sufficiently large energies, $\sigma_{ta\to dd}$ surpasses the elastic cross sections, thereby suggesting a similar behavior
for degenerate thresholds and not a convergence of reaction and elastic cross sections to the same value.
Having exhausted the $\aA$ alteration as another knob to explore sensitivity with respect to presumed unobservable short-distance structure,
it remains to change the \mbox{3-body} parameter directly, e.g., by using the second excited trimer as a scattering fragment.



\section{Summary}\label{sec.sum}
The \mbox{4-body} scattering system was analyzed in the framework of a non-relativistic, regularized, zero-range effective theory.
The threshold structure was chosen such that the usefulness of the framework could be assessed for a variety of 
reaction scenarios including nuclear fusion. Accordingly, the system of interest comprises an overall 
bosonic system -- two protons and two neutrons or any other \mbox{4-component-or} more-fermion system -- in the $J^\pi=0^+$ channel renormalized to a 
finite 2-body scale. We reported the following:
\begin{enumerate}
\item 
As a consistency check, we added further numerical evidence for the usefulness of the zero-range theory for rearrangement reactions by
quantifying the cutoff-regulator and renormalization-condition invariance of the \mbox{4-body} bound-state spectrum and
elastic and inelastic scattering amplitudes.
For a dimer-dimer to trimer-atom reaction, in particular, we considered the scattering 
system with excited states as \mbox{3-body} fragments. The results shown are insensitive to whether the
atom scatters off an exctited or the ground-state trimer as long as its energy is kept fixed.
The thereby installed renormalization of the \mbox{3-body} system not only comprises a more comprehensive scheme but,
through its avoidance of large coupling constants, a more practical one which more closely resembles the approach via
Faddeev(-Yakubovsky)-type integral equations.

\item Besides threshold effects, our variation of the gap between dimer-dimer and trimer-atom (i.e., deuteron-deuteron and 
triton-proton, in our nuclear interpretation) did not reveal the presence of potential \mbox{4-body} resonances or any other poles 
close enough to the scattering thresholds to have a significant impact.

\item The dominance of the reaction rate for dimer-dimer to atom-trimer
compared with the elastic rates when the respective thresholds are degenerate (i.e., 2 times dimer binding equals trimer binding 
energy) was found renormalization-group invariant.
This enhancement of the reaction probability contrasts an earlier study~\cite{Deltuva:2011ur}, indicating a strong sensitivity of our 
results for the \mbox{4-body} reaction rates to the details of the trimer wave function.
\end{enumerate}

In regards to that last result, we varied this ``interior" region of the interaction potential {\it via}
the cutoff regulator, through the renormalization condition, and both measures do not resolve the discrepancy: 
the reaction probability remains dominant.
Hence, we must conclude that these measures are an incomplete assessment of how the reaction cross sections respond to short-distance interaction 
structures. We showed that the ``distance'' of the 2-body system to the unitary limit strongly affects the ratio
between elastic and rearrangement scattering, thus providing a potential resolution to the
discrepancy of our results with other simulations. 
In conclusion, we stress the importance of assessing
the reaction rates further on their sensitivity
to exactly how the systems are moved away from the unitary 2-body limit. For nuclei, in particular,
(iso)spin dependent interactions can be considered at leading order. This allows, in principle, for more
rearrangement channels of which the singlet-dimer-singlet-dimer ones are unphysical and provide yet another
puzzle that awaits "physicalization" at higher orders in perturbation theory of the pionless EFT.
Furthermore, the ratio at low energies $\lesssim10~$MeV of reaction vs. elastic-scattering cross sections
which, as calculated with numerous high-precision nuclear-interaction potentials
(see compilations~\cite{Fonseca:2017koi,Lazauskas:2019hil}), exhibits an order-of-magnitude suppression of
the reaction cross section when thresholds are at the nuclear-physical locations. The values obtained here for
the cross sections for $\bd<B_{\tiny ^2\text{H}}$ and $\bt{0}<B_{\tiny ^3\text{H(e)}}$ do not exhibit this
separation of scales. Hence, the cross-section ratio represents an uncommonly strong sensitivity of an observable,
presumably within the range of applicability of the pionless EFT, with respect to changes in the imposed renormalization
constraints.

Finally, we motivate an investigation into how the number of \mbox{4-body} bound states depends
on characteristics of the \mbox{3-body} spectrum, namely, number of bound trimers, and location relative
to the dimer thresholds, with its potential universal value unique to four bosons.

\section*{Acknowledgments}
SM acknowledges financial support from the Department of Science and Technology (DST) under the scheme of INSPIRE (Fellowship 
number IF190758). UR acknowledges financial (research and travel) support from MATRICS and Core Research Grants from the Science 
and Engineering Research Board (SERB), Govt. of India (grant numbers MTR/2022/000067 and CRG/2022/000027). 
SM would like to thank Abhik Sarkar and Swarup K. Sarkar for their invaluable assistance in various numerical discussions. 
JK gratefully acknowledges hospitality and support of the
Department of Physics, IIT Guwahati, where part of this work was done.
Furthermore, JK is indebted to M.~Birse, L.~Contessi, R. Lazauskas, and A.~Deltuva for helpful comments. 

\appendix

\section{Regularization and Renormalization}
\label{app.reg}
We renormalize/calibrate the threshold structure of the low-energy 4-boson system {\it via} (a) employing a computationally 
convenient Gaussian model for the interactions in Eq.~\eqref{eq.ham0}, namely, with a 2-body potential given by
\begin{equation}\label{eq.v2}
\hat{V}_2 (r_{ij},\lambda) = c(\lambda)\,e^{-\lambda |r_i - r_j|^2}\,,
\end{equation}
and a \mbox{3-body} potential given by
\begin{equation}\label{eq.v3}
\hat{V}_3(r_{ij},r_{ik},\lambda) = d(\lambda)\,e^{-\lambda (|r_i - r_j|^2 + |r_i - r_k|^2)}\,,
\end{equation}
both with a zero-range limit for the regulator \( \lambda\to\infty \); and (b) fitting the cutoff dependencies of the strengths
of the potentials $c(\lambda)$ and $d(\lambda)$, such that:
\begin{gather}
\bd=0.5~\text{MeV}\approx 0.0025\,\text{fm}^{-1}\,,\\
1.2\,\bd\leq \bt{i}\leq 9\,\bd\quad\text{for}~i=0,1,2,\,\, \text{cf. Table~\ref{tab.a}}\,,
\end{gather}
In other words, each pair $(i,\lambda)$ represents an interaction for which we solve the scattering problem for the given range 
of threshold structures as set by $\bt{i}$. To avoid confusion, one of our calculations to assess the correlation of $\bd$ and,
e.g., $\bt{2}=1.5\bd$, with elastic dimer-dimer and trimer-atom scattering and the transition between those two arrangements, 
obtains $c(\lambda)$ to get $\bd$ and $d(\lambda)$ to yield $\bt{2}$ for cutoffs as shown in Table~\ref{tab.a}.

The choice to fix the dimer binding energy instead of the correlated 2-body scattering length is na\"ively expected to produce 
less variation of the respective scattering-fragment wave functions at distances large compared with the interaction's range. 
Nonetheless, if the latter procedure is chosen, an analytical calculation yields identical $\lambda$ dependencies $c(\lambda)$ 
for $\lambda\to\infty$. In the latter limit, the interaction Eq.~$\eqref{eq.ham0}$ cannot sustain a single \mbox{3-body} bound-state,
regardless of what one uses as $d(\lambda)$. Forcing the ground state to represent the trimer is
thus an approach to be abandoned beyond a certain critical cutoff. Neither in this work nor in many earlier
few-body calculations with coordinate-space-regulated interactions of type Eqs.~\eqref{eq.v2}~and~\eqref{eq.v3}~$\lambda$
dependence is probed close to that limit. Nevertheless, although all calculations in this work could
have been carried out with a ground-state condition, the significant difference of  the structure of the \mbox{3-body} potential
and the conjectured independence of \mbox{4-body} reaction with respect to this alteration motivates its usage.

\section{Variational method}
\label{app.rgm}
We variationally solve the scattering (Kohn-Hulth\'ene functional) and bound-state (Raleigh-Ritz 
functional) problem for $A\leq 4$ particles. For the former, the variational space is spanned by two types of basis vectors: 
one representing two non-interacting, asymptotically free fragments (either two dimers or a trimer plus an atom) moving
relative to each other in a non-normalizable scattering state; and the other type with non-zero support only when the four 
particles are close to each other, i.e., in a small hyperradius ($\rho$) configuration. The complete asymptotic scattering 
solutions which are fully specified by a boundary condition (bc), as adopted from Ref.~\cite{hmh} to yield a symmetric S-matrix,
can be expressed in the general form
\begin{equation}
\label{eq.totwfkt}
\psi_{\text{bc}}=\asy\!\left[\!\sum_{{k\in\atop \text{\tiny phys.\,ch.}}}\!\!\!\left(f_{k}\,\delta_{\text{bc},k}\!+\!g_{k} A_{\text{bc},k}\right)
\,+\!\!\sum_{{m\in\atop \text{\tiny dist.\,ch.}}} \!\!\! d_{\text{bc},m}\,\chi_{m}\right]\,,
\end{equation}
where $f_k$ and $g_k$ are the regular and regularized irregular parts of the full scattering solution with the first sum $k$ taken 
over all possible physical channels. The variational parameters $A_{\text{bc},k}$ describe the phase by which the full state is 
shifted in the asymptotic regime from a scattering solution without inter-fragment interactions. The parameters $d_{\text{bc},m}$ 
allow the full state's distortion when the fragments are close enough to interact. The corresponding distortion wavefunctions 
constitute a suitable number of square-integrable functions $\chi_m$ (such as Gaussians) which are so chosen to provide improved 
convergence of the solutions in the non-asymptotic regime.

In the former (scattering) problem, the physical channels comprise fragments in stable bound-states, namely, the dimer ground state 
and the first and second excited trimer states. As no long-range interaction is relevant at LO, the relative motion is given by 
(ir)regular spherical Bessel functions $(G_l)F_l$. A physical channel for our \mbox{4-particle} system is defined by the partition of the 
particles into two fragments, namely, dimer-dimer ($dd$) or trimer-atom ($ta$), the relative angular momentum $l$ of the 
inter-fragment motion, and the channel spin $J_c$ which couples the total angular momenta of the fragments ($[j_1\otimes j_2]^{J_c}$).
For practical reasons, we choose a convenient form of the scattering solutions $f_k$ and $g_k$, namely,
\begin{gather}
\label{eq.wfktfull}
\left\{f_k(\vec{r}),g_k(\vec{r})\right\} =\psi^J_{(j_1,j_2)J_c,l}(\vec{r})\,\left\{F_{l}(r),\tilde{G}_{l}(r)\right\}\,,
\nonumber\\
\text{with}\quad\psi^J_{(j_1,j_2)J_c,l}(\vec{r})=\left[\frac{1}{r}Y_l(\hat{\vec{r}})\otimes 
\left[ \phi_1^{j_1}\otimes \phi_2^{j_2}\right]^{J_c}\right]^{J}\,,
\end{gather}
where the angular part of the relative motion is absorbed in the channel function $\psi$. The fragment states have the structure
\begin{eqnarray}\label{eq.fragwfkt}
\phi^{j_1} &=&
\sum_{n}\Bigg[ c_n\bigg[\prod_{i=1}^{n(f)-1}e^{-\omega_{in}\vec{\rho}_i^2}\,\mathcal{Y}_{l_{in}}(\vec{\rho}_i)\bigg]^{L_n}
\nonumber\\
&&\otimes \,\, \asy\Big[\ket{s_{1}}\ldots\ket{s_{n(f)}}\Big]^S\Bigg]^{j_1}=:\sum_nc_n\braket{\tilde{\vec{\rho}}}{n}\,,
\nonumber\\
\end{eqnarray}
with $n(f)-1=1(2)$ Jacobi coordinates $\vec{\rho}$ describing the spatial motion within the dimer(trimer) and the tilde being shorthand
for a vector collecting both spatial and internal coordinates appropriately. Hereby, we expand a state with total spin $j_1$ as a product
of solid-harmonics ($\mathcal{Y}_l$) and Gaussians allowing, in principle, for any set of angular momenta ($l_n$) that contributes to the 
state of interest. The $c$'s are a solution of the generalized eigenvalue problem \mbox{$\sum_m(\mel{n}{\ham}{m}-e\braket{n}{m})c_m=0$} 
that expand the desired fragment bound-state wavefunction. The vector-coupling coefficients are implicit {\it via} the bracket notation. 
The structure of $\ham$ motivates a pure \mbox{$S$-wave} approximation according to which we set \mbox{$l_{in}=0$}, and hence, hereon it is understood
that \mbox{$(G)F:=(G_{l=0})F_{l=0}$.} Since this analysis is extended to include (iso)spin-dependent terms for a more realistic description of 
\mbox{4-component-fermion} systems such as nucleons, we label the internal spin ($\sigma,m_\sigma$) and isospin ($\tau,m_\tau$) states of a 
particle by the collective quantum number \mbox{$s_{i}\in\lbrace \sigma, m_\sigma, \tau, m_\tau\rbrace$} corresponding to the fermions belonging 
to a given fragment ($f$) with the $1\leq i\leq n(f)-1$. The overall bosonic character of our system is then enforced by antisymmetrizing 
(denoted by the operator $\asy$) the internal state, Eq.~\eqref{eq.fragwfkt}, even before $\asy$ acts on the total wave function, 
Eq.~\eqref{eq.totwfkt}.

The width parameters $\omega_{in}$ determine the spatial extent of the Gaussian functions used to expand the radial dependence. 
Specifically, we found numerically converged results in the range $\omega\in(10^{-4},40)\,\text{fm}^{-2}$. While, for the dimer fragment
we obtained the ground-state energy corresponding to the renormalization condition with 5 to 6 widths, for the trimer fragment about 
50 to 70 parameter combinations for the two Jacobi coordinates were necessary. For a complete expansion of the \mbox{4-body} state, it is also 
convenient to express the non-normalizable inter-fragment (relative) motion function $F_l$ in Eq.~\eqref{eq.wfktfull} in terms of Gaussian
functions. Bound-state and scattering observables were found converged by using about 25 to 30 parameters whose magnitudes, however, differ
significantly from those used for the individual fragment bound-states as higher precision is usually needed to expand the (ir)regular 
scattering solutions of the free inter-fragment motion in the limit of zero momentum. While this zero-momentum character of the function 
demands a wide range of basis functions, the regularization of the irregular solution requires quite the opposite to avoid an unphysical 
dependence on this numerical regulator.
Specifically, to ensure that the irregular solution \( G(r)\propto r^{-1}\,\sin(kr) \) remains well-behaved at \( r = 0 \), we instead use 
in Eq.~\eqref{eq.wfktfull} the regulated version
\begin{equation}
\tilde{G}(r) = \beta r\,e^{-\beta r} G(r)\,.
\end{equation}
No observable should depend on the regularization parameter $\beta$ which, na\"ivley, is expected if it alters the relative-motion function
only at distances where the inter-fragment interaction is non-zero, and hence, assumes a form different compared with the asymptotic solution. 
As our EFT potentials cover a range of support, and because it is the potential effective between the fragments, we validated the 
$\beta$-independence of our results for each EFT-regulator value $\lambda$ and \mbox{3-body} parameter, separately.

The Gaussian basis is naturally limited in its ability to expand non-normalizable functions. This does not of course render the approach 
useless as far as the solution to the variational problem, Eq.~\eqref{eq.kohn}, is concerned, given that the only matrix elements between 
those asymptotically non-vanishing functions contribute whose kernel goes to zero with the inter-fragment interaction. Hence, such free 
functions need to be expanded accurately only up to a certain distance beyond which the matrix elements are oblivious to the functions.
Given a set of width parameters which is appropriate for an expansion of the relative motion up to that distance, we employ a standard, 
weighted minimization of the form
\begin{equation}
\int_{0}^{\infty} \left(F(r) - \sum_{m} a_{m} \chi_{m}(r)\right)^2 W_\epsilon(r) \, dr\,,
\end{equation}
with a weight function
\begin{equation}
W_\epsilon(r) = e^{-\epsilon r^2}/r\,.
\end{equation}
The larger the value of \( \epsilon \), the better is the Gaussian expansion for small distances, while for $\epsilon=0$ all points are 
weighed equal and an expansion of the free wave up to infinity is sought after. In our calculations, we assessed the result's sensitivity 
for \( \fm{10^{-4}}<\epsilon<\fm{10^{-2}} \). At both ends, results become highly sensitive to $\epsilon$: if chosen too small, the fit
is forced to sacrifice accuracy at short distances in order to better expand the irrelevant long-distance parts; if $\epsilon$ is too large,
the relative wave is not well described where it still contributes to variational integrals. Within 
the above-stated interval, however, we find a stable plateau where the expansion is appropriate.

Having thus defined the variational space fully, allows us to 
solve Eq.~\eqref{eq.kohn}.
From a ``Kato-corrected''~ \cite{kato1951upper}~reactance matrix $A_{\text{\tiny ch,ch'}}$,
we obtain the scattering matrix 
\begin{equation}
S = \frac{1 + i A}{1 - i A}\quad,
\end{equation}
whose elements are parametrized with an
inelasticity factor \( \eta \) and a real phase shift \( \delta(E) \) 
(see e.g.~\cite{newton2002scattering}). The element characterizing the
transition $\text{ch}\to\text{ch'}$ reads
\begin{equation}
S_{\text{\tiny ch,ch'}} = \eta_{\text{\tiny ch,ch'}} \, e^{2i \delta_{\text{\tiny ch,ch'}}(E)}\quad.
\end{equation}
Hence, the cross sections calculated via~Eq.~\eqref{eq.trans}~are functions solely of $\eta$.

\section{Genetic Algorithm for Parameter Optimization}{\label{GA}}
Variational parameters determined from either functional Eq.~\eqref{eq.ritz}~and Eq.~\eqref{eq.kohn}~represent
linear superposition coefficients for functions which are assumed to span the space of bound-state functions
(dimer and trimer fragments) and \mbox{4-body} scattering states completely. Variation does not entail the
non-linear width parameters and the (iso)spin and orbital-angular-momentum coupling schemes.
While the latter set is limited to the pure $S$-wave sector due to the interaction character and our
focus on low-energy amplitudes, the selection of Gaussians demands numerical optimization.
We ensure an appropriate representation of the scattering fragments by optimizing
the basis-vector characterizing widths in Eq.~\eqref{eq.fragwfkt} via a genetic evolution
(see e.g.~\cite{Davis1990HandbookOG}).
It noteworthy to stress that optimizing a basis such that it spans a space within which the
dimer and trimer are bound by the energies the Hamiltonian was renormalized with may
introduce significant model dependence. While the binding energy might not depend on
whether Gaussians -- as done here -- or, e.g., Lorentzians are used, the corresponding ground-state
wave functions might. To us, it is not obvious how this potential difference becomes part of the
short-distance sensitivity which we quantify with the EFT-cutoff $\lambda$ variation -- we \emph{assume} it does.
Under this assumption, seek bases with a relatively small condition number a large number of eigenvalues
in an energy region including thresholds; e.g. the accuracy of a \mbox{4-body} basis in expanding
scattering of an atom off the second excited state of a trimer is assumed to increase with the density of
its eigenvalues in the vicinity of $\bt{2}$. 
We drive the evolution towards such bases with the fitness function
\begin{gather}\label{eq.fitness}
\mathcal{F}=
    \frac{\sum_{i\in \mathcal{S}} e^{-0.007 \cdot e_i}}{\max\{e_i\}} \cdot \sqrt{\log(1.1 + N_<)}
    \nonumber\\ 
    \text{if} \quad \ C > C_{\text{min}}\sim10^{-12} \quad \text{and} \quad 0 \quad \text{otherwise.}
\end{gather}
    \( e_i \) is the $i$-th Hamiltonian eigenvalue value, \( C \) is the condition number (ration between
    smallest to largest norm eigenvalue), and \( N_< \) is the number of eigenvalues below a threshold we
    chose for each $\lambda$ and \mbox{3-body} scenario according to Table~\ref{tab.spect}~such that the threshold of
    interest is included. As the excited-state wave functions of the trimer targets differ significantly in
    shape from those of ground states, the fitness is measured with respect to the set of eigenvalues $\mathcal{S}$.
    If the ground state should be optimized, $\mathcal{S}={0}$, if the first two excited states are of interest, 
    $\mathcal{S}={1,2}$. While the latter choice yielded also well converged ground-state energies, the former
    set's excited states could differ by $>10\%$. The denominator and the ostensibly random factor
    account for the difference in magnitude of the considered numbers such that changes in both $N_<=\mathcal{O}(1)$
    and $e^{-\#\,e_i}$ affect $\mathcal{F}$ by the same amount. The algorithm follows the canonical steps:
    \begin{enumerate}
        \item \textit{Initialization} of a randomly chosen set of bases (seed population);
        \item \textit{Fitness Evaluation} based on the Hamiltonian spectrum (Eq.~\eqref{eq.fitness};
        \item \textit{Selection} of a subset of the population (parents) with preference to fitter individuals;
        \item \textit{Crossover} of binary representations of parents breeding correlated bases (offspring) including
        a random mutation probability (bit flip) of $0.2\%$;
        \item \textit{Insertion} of fit-enough offspring into the population and subsequent removal of the lowest-ranked
        individuals thus keeping the population's size constant (in practise, we updated between 30 and 60 bases
        in parallel while replacing 6 to 8 members per generation);
        \item \textit{Iteration} over multiple (typically 20 to 40) generations until no significant change in fitness
        and stability is observed;
    \end{enumerate}
In addition to optimizing fragment functions, we employed this algorithm for the much larger \mbox{4-body} bound-state
bases which we added as distortion channels (the $\chi$'s in Eq.~\eqref{eq.totwfkt}) in order to allow maximal
variational freedom in the wave function when fragments are close and are not simple products of non-interacting
stable states.

\bibliography{4bdyCCscatt.bib}

\end{document}

%% file: graphs/4bdyEfimovGraph.tex
\begin{tikzpicture}[x=1cm,y=1cm,font=\scriptsize,scale=1.5]

\definecolor{green}{rgb}{0.0,0.5,0.0}

\tzaxisx[gray,thick]{-3}{5}{\large $a_2^{-1}$}[black]
\tzaxisy[gray,thick]{-5}{.25}{\large $E$}[black]

\tzaxisy[-,red,thick]<2>"vert"{-5}{0.25}{\small $a_2\approx 10~\textrm{fm}$}[red]

\def\ddx{-0.5*(\x)^2}
\def\daax{-0.25*(\x)^2}

\tzfn[gray,thick]\daax[0:4]{\small $daa$}[r]
\tzfn[black,thick]\ddx[0:3.1]{\small $dd$}[r]

\tzvXpointat{ddx}{2}(K1)
\tzvXpointat{daax}{2}(K2)
\tzgetxyval(K1){\Kx}{\Ky}
\tzgetxyval(K2){\kx}{\ky}

\tzhXpointat{daax}{1.4*\ky}(K3)
\tzgetxyval(K3){\KxII}{\KyII}

\tzhXpointat{daax}{\Ky}(K4)
\tzgetxyval(K4){\KxIV}{\KyIV}

\tzhXpointat{daax}{1.4*\Ky}(K5)
\tzgetxyval(K5){\KxV}{\KyV}


\tzhfnat[green,left,thick]{\ky}[\kx:0]

\node[anchor=south] at (-1,0) (aI) {I};
\node[anchor=west] at (0,\ky) (eI) {};
\node[green,right= of eI,xshift=-55pt,yshift=7pt] (i) {\large $B_2$};
\draw[green, rounded corners=15pt,thick] (aI) |- (eI);

\tzto[dashed,green,out=-0,in=170,thick]"curve"(2,\ky)(5,1.2*\ky){\small{$t^{(2)}_{\textrm{\tiny I}}\!\!-a$}}[r]

\node[green] at (1.5*\kx,0.5*\ky) (tI) {$B^{(2,\textrm{shallow})}_4$};
\draw[green,->] (tI) -- (1.2*\kx,1.05*\ky);

\tzhfnat[green,left,dotted,very thick]{1.05*\ky}[0.8*\kx:1.2*\kx]

\node[anchor=west] at (0,-4.5) (eI) {};

\node[green,right= of eI,xshift=-20pt,yshift=-12pt] (i) {\large $9\cdot B_2$};
\draw[green,->] (i) -- (\Kx,-4.5);

\tzto[dashed,green,out=-0,in=-180,thick]"curve"(0,-4.1)(\Kx,-4.5){}[r]
\tzto[dashed,green,out=180,in=0,thick]"curve"(0,-4.1)(-3,-3.2){}[r]
\tzto[dashed,green,out=-0,in=165,thick]"curve"(2,-4.5)(5,-4.9){\small{$t_{\textrm{\tiny I}}^{(1)}\!\!-a$}}[r]

\tzhfnat[orange,left,thick]{1.4*\ky}[0:\KxII]

\node[anchor=south] at (-1.6,0) (aI) {II};
\node[anchor=west] at (0,1.4*\ky) (eI) {};
\draw[orange, rounded corners=15pt,thick] (aI) |- (eI);

\tzto[dashed,orange,out=-0,in=170,thick]"curve"(\KxII,1.4*\ky)(5,1.7*\ky){\small{$t_{\textrm{\tiny II}}^{(1)}\!\!-a$}}[r]

\tzhfnat[orange,left,dotted,very thick]{1.65*\ky}[0.8*\kx:1.2*\kx]

\tzhfnat[orange,left,thick]{1.4*\Ky}[0:\KxV]
\tzto[orange,out=-0,in=180,thick]"curve"(-3,1.4*\Ky)(0,1.4*\Ky)
\tzto[dashed,orange,out=-0,in=160,thick]"curve"(\KxV,1.4*\Ky)(5,1.7*\Ky){\small{$t_{\textrm{\tiny III}}^{(1)}\!\!-a$}}[r]

\tzhfnat[blue,left,thick]{\Ky}[0:\KxIV]
\node[anchor=south] at (-2.5,0) (aI) {IV};
\node[anchor=west] at (0,\Ky) (eI) {};
\node[blue,right= of eI,xshift=-70pt,yshift=7pt] (i) {\large $2\cdot B_2$};
\draw[blue, rounded corners=15pt,thick] (aI) |- (eI);

\tzto[dashed,blue,out=-0,in=165,thick]"curve"(\KxIV,\Ky)(5,1.2*\Ky){\small{$t_{\textrm{\tiny IV}}^{(1)}\!\!-a$}}[r]

\node[green] at (0.5*\kx,1.3*\Ky) (tI) {\tiny $B^{(2,\textrm{deep})}_4\approx \textcolor{blue}{B^{(1,\textrm{shallow})}_4}$};
\draw[green,->] (tI) -- (0.8*\kx,1.03*\Ky);
\tzhfnat[green,left,dotted,very thick]{1.03*\Ky}[0.8*\kx:\kx]{}[l]

\tzhfnat[blue,left,dotted,very thick]{1.03*\Ky}[\kx:1.2*\kx]{}[l]

\tzcdot*[red] (K1)(1.5pt)
\tzcdot*[red] (K2)(1.5pt)


\node[orange] at (1.65*\kx,1.85*\Ky) (tI) {\scriptsize $B^{(1,\textrm{shallow})}_4$};
\draw[orange,->] (tI) -- (1.2*\kx,1.63*\Ky);
\draw[orange,->] (tI) -- (1.2*\kx,1.65*\ky);
\tzhfnat[orange,left,dotted,very thick]{1.63*\Ky}[0.8*\kx:1.2*\kx]{}[l]

\end{tikzpicture}